# JDATATRANS for Array Obfuscation in Java Source Code to Defeat Reverse Engineering from Decompiled Codes

Praveen Sivadasan
School of Computer Sciences
Mahatma Gandhi University
Kerala, India
+91-9946063774
praveen_sivadas@yahoo.com

P Sojan Lal
School of Computer Sciences
Mahatma Gandhi University
Kerala, India
sojanlal@gmail.com
India

Naveen Sivadasan
TCS Innovation Labs Hyderabad
Tata Consultancy Services
Madhapur, Hyderabad, India
s.naveen@atc.tcs.com

## ABSTRACT
*Software obfuscation or obscuring a software is an approach to defeat the practice of reverse engineering a software for using its functionality illegally in the development of another software. Java applications are more amenable to reverse engineering and re-engineering attacks through methods such as decompilation because Java class files store the program in a semi complied form called 'byte' codes. The existing obfuscation systems obfuscate the Java class files. Obfuscated source code produce obfuscated byte codes and hence two level obfuscation (source code and byte code level) of the program makes it more resilient to reverse engineering attacks. But source code obfuscation is much more difficult due to richer set of programming constructs and the scope of the different variables used in the program and only very little progress has been made on this front. Hence programmers resort to adhoc manual ways of obscuring their program which makes it difficult for its maintenance and usability. To address this issue partially, we developed a user friendly tool JDATATRANS to obfuscate Java source code by obscuring the array usages. Using various array restructuring techniques such as 'array splitting', 'array folding' and 'array flattening', in addition to constant hiding , our system obfuscate the input Java source code and produce an obfuscated Java source code that is functionally equivalent to the input program. We also perform a number of experiments to measure the potency, resilience and cost incurred by our tool.*

## 1.Introduction
The java based web applications gained popularity because of its Architecture Neutral Distribution Format (ANDF) [12]. During compilation, the Java source code is translated to java class files that contain Java Virtual Machine (JVM) code called the 'byte code', retaining most or all information present in the original source code [5]. This is because the translation to real machine instruction happens in the browser of the user's machine by JIT (Just-In-Time Compiler). Also, Java programs are small in size because of the vast functionalities provided by the Java standard libraries.

Decompilation is the process of generating source codes from machine codes or intermediate byte codes. JAD, Mocha, Decaf are some of the well-known decompilers [22]. Though decompilation is in general hard for most programming languages, the semi compiled nature of Java class files make it more amenable to reverse engineering and re-engineering attacks through decompilation [10, 14, 15, 16]. This makes it easier for the competitors to extract the proprietary algorithms and data structures from Java applications in order to incorporate them into their own programs in order to cut down their development time and cost. Such cases of intellectual property thefts [6, 18, 19] are difficult to detect and pursue legally. Recent statistics [7] show that four out of every ten software programs is pirated worldwide and over the years, global software piracy has increased by over 40% and has caused a loss of more than 11 billion USD [7]. Over the years, a number of software protection methods have been proposed. The remote service based methods provide the maximum protection against piracy because the application resides in a remote server and only the results of the computation is returned to the client application without exposing the algorithmic details of the server application. But such methods suffer from limited network bandwidth and latency. Even the approach of running only the crucial software components in a remote server suffer from similar drawbacks [5]. The approach of encrypting the executable is effective only if the entire decryption/execution process takes place in the hardware [5]. Furthermore, there is dramatic difference in the cost of encryption and decryption in any public key encryption system [5].

Software obfuscation [6,9,13,17,21] is a popular approach where the program is transformed into an obfuscated program using an 'obfuscator' in such a way that the functionality and the input/output behavior is preserved in the obfuscated program whereas it is much more difficult to reverse engineer the obfuscated program. Though, obfuscation is a more economical method for preventing reverse engineering[5], there are 'deobfucators' [23, 24] available to defeat some of the less sophisticated obfuscation strategies. The popular transformation techniques employed for obfuscation are (i) *layout transformation* which makes the structure of the transformed program difficult to comprehend (ii) *data transformation* that obscures the crucial data and data structures (iii) *control transformation* to obscure the flow of execution [5, 6, 25]. The obfuscation can be preformed on the source code [4, 5], the intermediate code or the machine executable code. The effectiveness of obfuscation is usually measured in terms of a) the *potency* that is the degree to which the reader is confused, b) the *resilience* that is the degree to which the obfuscation attacks are resisted and finally c) the *cost* which measures the amount of execution time/space penalty suffered by the program due to obfuscation [5, 26].

The existing Java byte code obfuscators are primarily based on lexical transformations, where the class names, variable names and function names are replaced by less comprehensible strings. In source code obfuscators, the commonly applied transformations are (i) replacing symbol names with non-meaningful ones, (ii) substitution of constant values with arithmetic expressions, (iii) removing source code formatting, and (iv) exploiting the preprocessor. We refer the reader to [4, 7, 27, 28, 29] for a survey of different java obfuscation tools that are available. Little progress has been made so far in successfully applying more sophisticated obfuscation strategies for either byte code or source code obfuscation. This is primarily because of the difficulty in handling issues such as the scope of the variable names, dynamic binding of variable names to objects, polymorphism etc.

Data transformation and constant hiding are the two well studied obfuscation techniques. In data transformation, Array transformation in particular is popular. Array splitting, array folding and array flattening are the three well known array transformation methods [20,30,31]. As shown in Figure 1, in array splitting, a one dimensional array 'A' for example is split into say 'k' arrays A1 ... Ak such that array Ai holds the elements of 'A' with indices (i mod k). In array folding, a one dimensional array 'D' is transformed into a multidimensional array say for example a two dimensional array 'D1' using a transformation operation, which is a bijection between the indices of D and D1. Array flattening on the other hand does the reverse where a multidimensional array is transformed into a single dimensional array using the bijection mapping. In [7], Ertaul et. al proposed a novel constant hiding techniques using *y-factors*. The y-factors are essentially a predefined increasing sequence of 'm' prime numbers y[0], y[1],y[2]...,y[m]. The y_factors can be used to transform a non negative number 'x' which is less than y[0] as follows. Let the function 'F(A, k)' be defined as F(A, k) = ((....((A mod y[k]) mod y[k-1]) mod y[k-2]) .... mod y[0]). Now replace 'x' by the expression F(A, k) such that F(A, k) evaluates to 'x'. Now to hide any large positive constant say 'c' in the program, first 'c' is replaced with a simple expression of the form 2*d + r where 'r' is 0 if 'x' is even and 'r' is 1 if 'x' is odd. Now, the constants 2 and ‘r’ in the resulting expression can be hidden by replacing it with the corresponding F( ) function.

## Our Contribution

We developed a source code obfuscation tool JDATATRANS that obfuscate arrays in Java source code. Our tool has the following two major components.

***JDATATRANS-CoBS*** *(Classes for oBfuscating Source codes)*

An extensible repository of array generic classes which we refer to as CoBS (Classes for oBfuscting Source codes). The internal implementation of the array class is highly obfuscated. At present the repository has three separate array implementations using the well known obfuscations methods - array folding, array flattening and array splitting. The programmer has the choice of using any of these array implementations for each of the crucial arrays in the program, by instantiating the array object to the respective CoBS array class.

***JDATATRANS-Obfuscator*** *(for obfuscating the CoBS arrays in the Java program)*

We have developed an obfuscator that identifies the usage of the CoBS arrays in the Java program and obfuscates the corresponding sentences hiding the constant and array indices in it using the F( ) functions. This provides an additional level of obfuscation to the program in addition to the obfuscation provided by the CoBS implementation.

Our source code obfuscator produces a functionally equivalent Java source program and additional levels of obfuscation can be obtained by applying any of the existing byte code obfuscators on the target class files. We also perform a number of experiments to evaluate the effectiveness of our obfuscation system. Our experiments reveal that the system is able to make the program sufficiently incomprehensible even for decomplier assisted reverse engineering without much overhead in terms of the increase in code size or execution time. To the best of our knowledge, there are no existing array obfuscators for either Java source code or byte code. Furthermore, we believe that our approach of developing a library of highly obfuscated data structures as CoBS classes together with an obfuscator that reads the program and obfuscates the statements where CoBS objects are accessed is novel and is a significant step towards building high quality obfuscators for Java applications.

## 2.Implementation

In this section we give an overview of the various implementation aspects of JDATATRANS. As mentioned in the introduction, the two major components of the JDATATRANS tool are a) *JDATATRANS-CoBS* (Classes for oBfuscating Source codes) repository that contain generic classes for various obfuscated array implementations and b) *JDATATRANS-obfuscator* that obfuscates the Java programs that uses the CoBS arrays. Before we discuss the implementation details of these two components, we give an outline of the *ConstHide* module that hides the constants in the source code. The ConstHide module is used by both CoBS and the Obfuscator. The tool is built using Java 5.0 with user friendly GUI support based on SWING.

*Array Splitting: One dimensional array A is split into A1 and A2.*

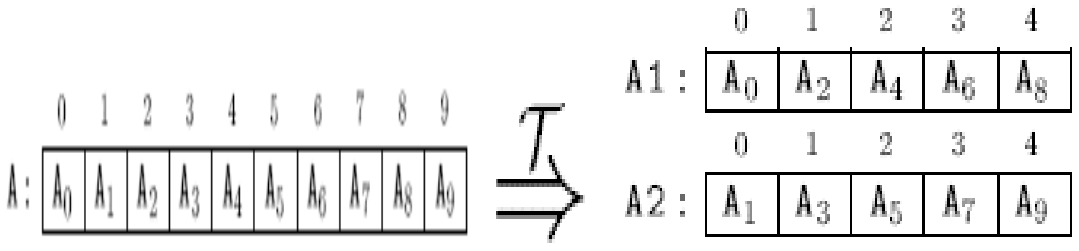

```
(1) int A[10];       (1) int A1[5], A2[5] ;
(2) A[i]=…;          (2)  if ((i%2)==0)
                            A1 [i/2] =…….;
                         else
                            A2 [i/2] =…….;
```

*Array Folding: One dimensional array D is folded into a two dimensional array D1.*

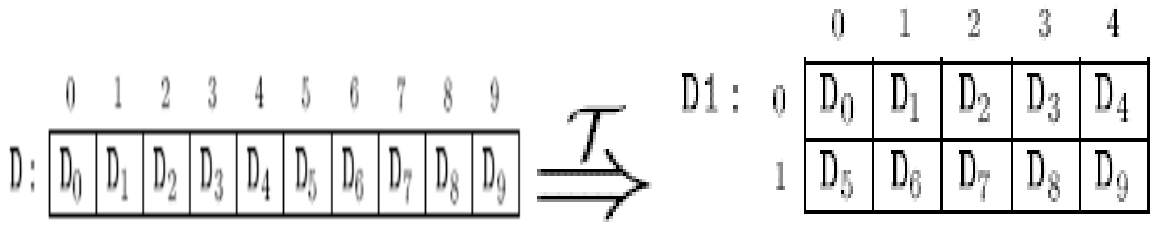

```
(1) int D[10];  (1) int D1[2][5];
(2) D[0]=…;      (2) D1[0][0]=……;
(3) D[5]=….;    (3) D1[1][0]=……;
(4) D[i]=….;    (4) D1[(i-(i%5))/5][i%5]=……;
```

*Array flattening: Two dimensional array E is flattened it into a one dimensional array E1.*

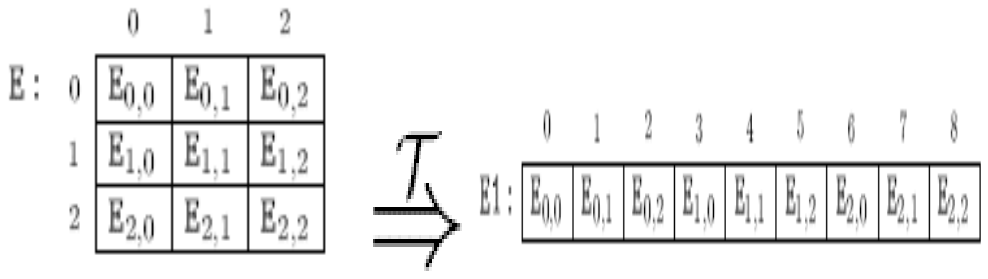

```
(1) int E[3][3];                 (1) int E1[9];
(2) E[i][j]=…;          (2) E1[3*i+j]=……;
```

Figure 1. The array restructuring techniques – array splitting, array folding and array flattening.

## 2.1 The ConstHide Module

To compute F( ), we use an array Y[m] of 'm' pairs where Y[i] = (Pi, Qi) denote the pair at the i-th index of Y. These pairs have the following property

a) for any pair Y[i] = (Pi, Qi), Pi + Qi is a prime number and b) if i < j then Pi + Qi < Pj + Qj. That is, sum of the numbers in any pair is a prime number and the pairs are stored in Y array in the increasing order of their sum value. The following sequence of pairs for example can be the contents of the Y-array - (2,3),(5,6),(11,12),(23,24),(47,48),(95,96),(191,192),….. (12287, 12288).

Following is the algorithm to compute F( ) function.

```
int F(A, k){
    //k is a number between 1 and m which
    //denotes the depth of the obfuscation.
     Y[m]={(P1,Q1),(P2,Q2)........(Pm,Qm)}
     r = A;
     for (i :k .....1) {r = r mod (Pi + Qi);}
     return r; }
```

The input constant in transformed to a corresponding F( ) expression using the following hide( ) function.

```
String  Hide (c) {
    //This function returns an expression
    //of the form F(A, k) which evaluates to
    // c

     Let c = 2d + r;
    //Note that r = 0 if c is even, else
    // r = 1
    //Now we will hide the first integer
    //  2 in the above expression

    Choose k randomly from {1 ... m}
    Let A be such that 2 = F (A, k)
    Choose two number B and C such that
    A=B mod C

    return the expression
          F(B mod C, k)*d + r;
  }
```

The ConstHide would for example hide the constant '2' by replacing 2 with any of the following expressions: F(41%23,2), F(374%191,5), F(757%383,6). Though most compilers simplify the expressions of the form 374%191, we still use these expressions to ensure that the source code itself is difficult to comprehend.

## 2.2JDATATRANS-CoBS

The repository at present holds the generic class implementations for split arrays, folded arrays and flattened arrays. The following shows the public methods for split arrays.

```
public class  SplitArray <E>{
  public  SplitArray (int size );
  public void setArray(int pos, E elem):
  public  E getArray(int pos);
  public int lengthArray();
}
```

The methods 'setArray( )', 'getArray( )' and 'lengthArray( )' are used to store an element at a given index, retrieve the element stored at the given index and to get the length of the array respectively. Implementation for these three methods is mandatory for all the array classes in CoBS repository. If the programmer wishes to use an obfuscated array for one of the crucial arrays say 'X', then he/she first needs to decide which obfuscation technique to use (say the splitarray mechanism is chosen) and simply need to include the following array declaration statement in the program for 'X' (Assume that X is an array of type Integer and size 1000).

```
SplitArray<Integer> X =
          new SplitArray <Integer>(1000);
```

When the programmer imports the CoBS package, initially the CoBS arrays have only dummy (stub) implementation for all the public methods. This is done for the following reason. First, it ensures that the program that uses the CoBS arrays compile. Now if there are multiple implementations of the SplitArray (with differing internal implementations) itself in the CoBS repository, The CoBS handler in the next phase, replaces the dummy implementation with one from the available implementations in a random fashion. This ensures an additional level of obfuscation.

The following is a sample obfuscated implementation of getArray( ), setArray( ) and lengthArray( ) for SplitArrays.

```
public class  SplitArray <E> extends obfuscate
```

```
{   E[] iObj1;E[] iObj2;
  public  SplitArray (int size )
  {  if((size% F(41%23,2))==0)
     {
       iObj1=(E[])newObject[(int)(size/F(1524%767,7))];
       iObj2 =(E[])newObject[(int)(size/F(88% 47,3))];}
    else {
        int temp=(int)(size/2)+1;
        iObj1= (E[])new Object[temp];
        iObj2= (E[])new Object[size-temp];}
  }
  public void setArray(int pos,E elem){
       if((pos% F(183%95,4))==0)
               iObj1[(int)pos/ F(374%191,5)]=elem;
    else
               iObj2[(int)pos/ F(757%383,6)]=elem;
  }
  public  E getArray(int pos){
          if((pos% F(1524%767,7))==0)
                return(iObj1[(int)pos/ F(3059%1535,8)]);
   else
               return(iObj2[(int)pos/ F(6130%3071,9)]);
  }
  public int lengthArray() {
        return(iObj1.length+iObj2.length);
  }
}
```

Note that all the constants are hidden using the F( ) functions returned by the ConstHide module. The internal implementation can be obfuscated to any level of sophistication. And as mentioned earlier, the system supports multiple implementations of Split arrays with varying levels of obfuscation and the CoBS handler make the choice when the programmer uses the CoBS classes.

## 2.3JDATATRANS-Obfuscator

The JDATATRANS-Obfuscator scans the program and identifies those statements, called candidate statements, in the program where either a CoBS based array is declared or an instance of the array is accessed using any of its public methods. To do this, the obfuscator first preprocesses the input program. In the preprocessing phase, the sentence boundaries are detected, the comments are stripped off and the tokens in the sentence are identified. Now in each candidate statement, the obfuscator identifies the constants used, including the array indices. If there are no constants used then the array index variable is muliplied by '1' (which clearly does not alter its value) and the constant '1' is hidden using ConstHide. If only the lengthArray( ) function is invoked, then it is similarly replaced by lengthArray( ) * 1, where the '1' is later hidden by ConstHide. For non candidate statements, the first integer constant is hidden, avoiding alphanumeric strings. We remark that the resulting program can be rebofuscated again by the obfuscator to obtain further levels of obfuscation.

To illustrate the method, consider the following snippet from the program 'myprog.java' which the programmer wishes to obfuscate. The programmer decides to obfuscate the array 'ar' using SplitArray.

```
SplitArray<Integer>ar=newSplitArray<Integer>
(100000);
ar.setArray(i,(3*i + 1000) % n);
y = ar.getArray(i);
```

After obfuscation, it is transformed into the following code.

```
SplitArray<Integer>ar=newSplitArray<Integer>
(50000*F(49135%24575,12));
ar.setArray(i*(4*F(3059%1535,8)(F(49135%24575,12)*
F(35%27,2)+F(33%21,2))),(3*i + 1000)% n);
y=ar.getArray(i*(F(35%27,2)-F(12273%6143,10)));
```

After one more iteration of obfuscation, it is further transformed into the following code.

```
SplitArray<Integer>ar=newSplitArray<Integer>
(50000*F((F(49135%24575,12)*24567+F(33%21,2))%2457
5,12));
ar.setArray(i*(4*F((F(49135%24575,12)*1529+F(33%21
,2))%1535,8)(F(49135%24575,12)*F(35%27,2)+F(33%21,
2))),(3*i + 1000) % n);
y=ar.getArray(i*(F((F(49135%24575,12)*17+F(33%21,2
))%27,2)-F(12273%6143,10)));
```

But this creates an additional overhead in terms of the execution time as the number of F( ) expressions that needs to be computed in runtime increases with each iteration of the obfuscation.

Figure 2 shows how the various JDATATRANS components that we discussed so far interact in order to obfuscate the input Java program.

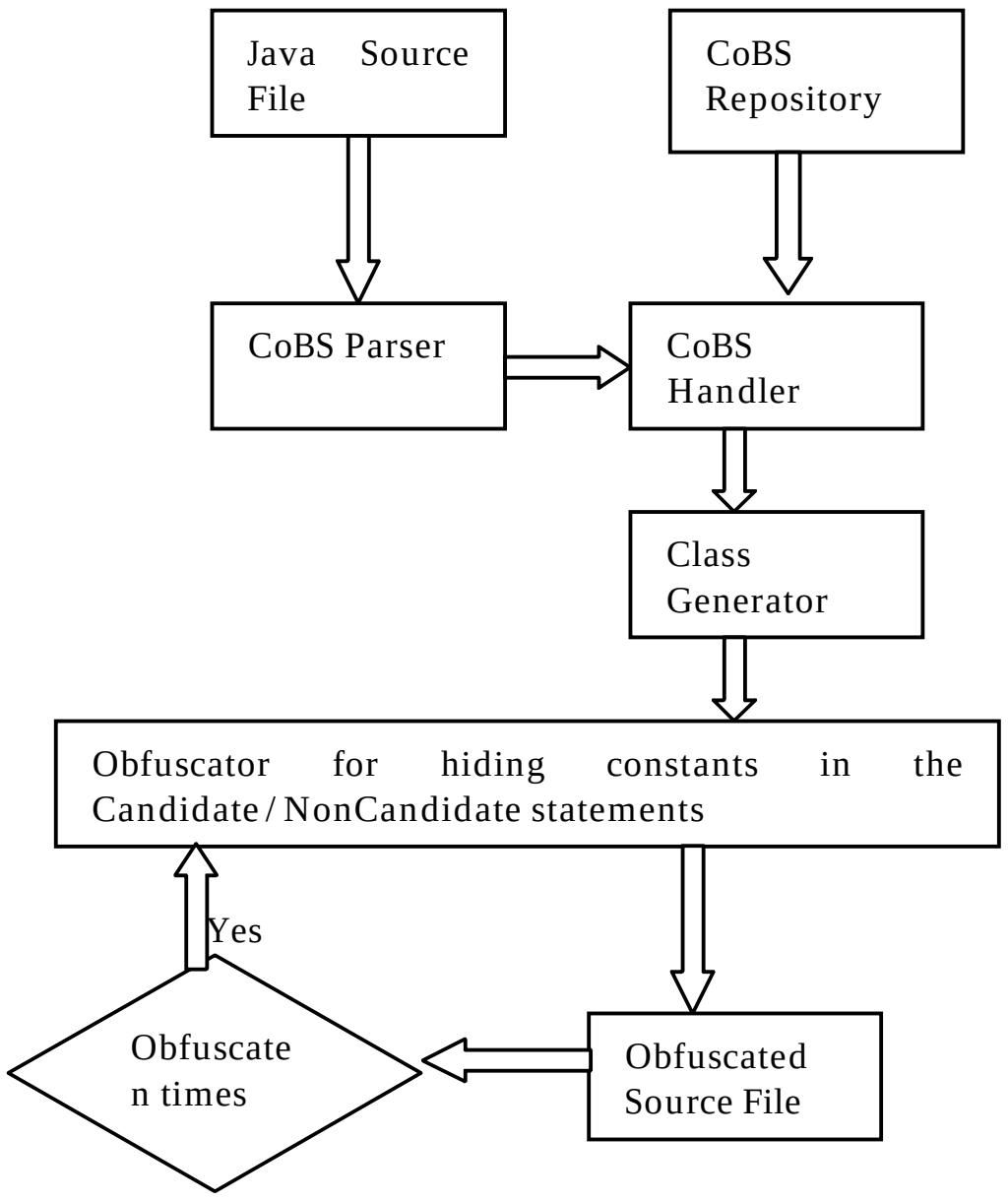

Figure 2. High-level view of the JDATATRANS components

## 3.Experimental Results

The following plot shows the tool performance where the analysis is performed on a sample code 'myprog.java' denoted by A and its obfscated version using SplitArray, FoldedArray and FlattenedArray denoted by B, C and D respectively. The algorithm section of ‘myprog.java’ is as follows

*Set 'n' elements to an array of size 100000*
*Print n array elements*

The execution time analysis is performed on a system with Intel Core Duo processor, 1.66GHz, with 1GB of RAM.

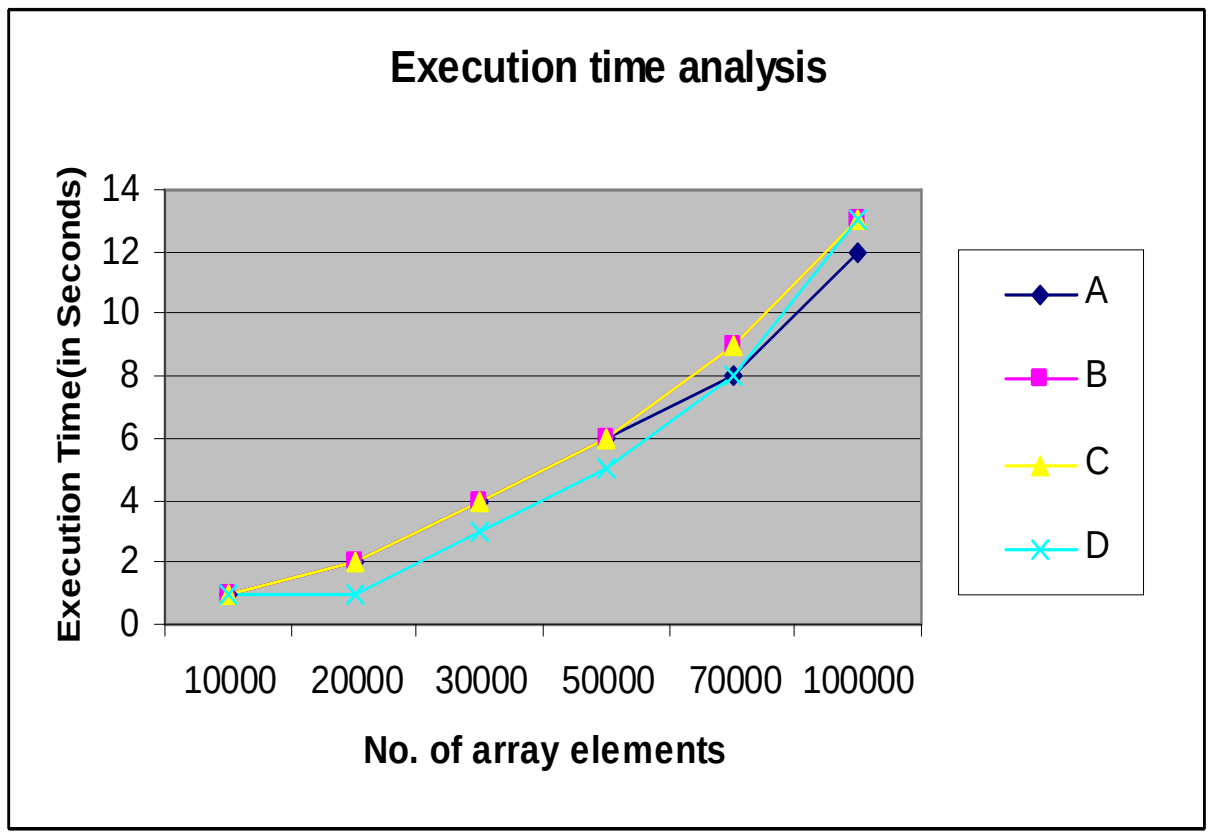

The graph shows no major variations in execution time for A, B, C, D for different number of array elements. For 100000 elements, the execution time analysis of A, B, C, D and its obfuscated codes are performed. Let P2, P3, P4, P5 correspond to different obfuscated versions of codes A, B, C, D. For code say, B (myprog_SplitArray.java), the obfuscated versions are B2 (myprog_SplitArray_mod123.java),B3(myprog_SplitArray_mod 123123.java),B4(myprog_SplitArray_mod123123123.java),B5(myprog_SplitArray_mod123123123123.java).The tool output B1 (myprog_SplitArray_mod.java) is a formatted nonobfuscated version of B. Let P,P1,P2,P3,P4,P5 corresponding to codes A,B,C and D be represented on X-axis and the Execution time(Sec) on Y axis.

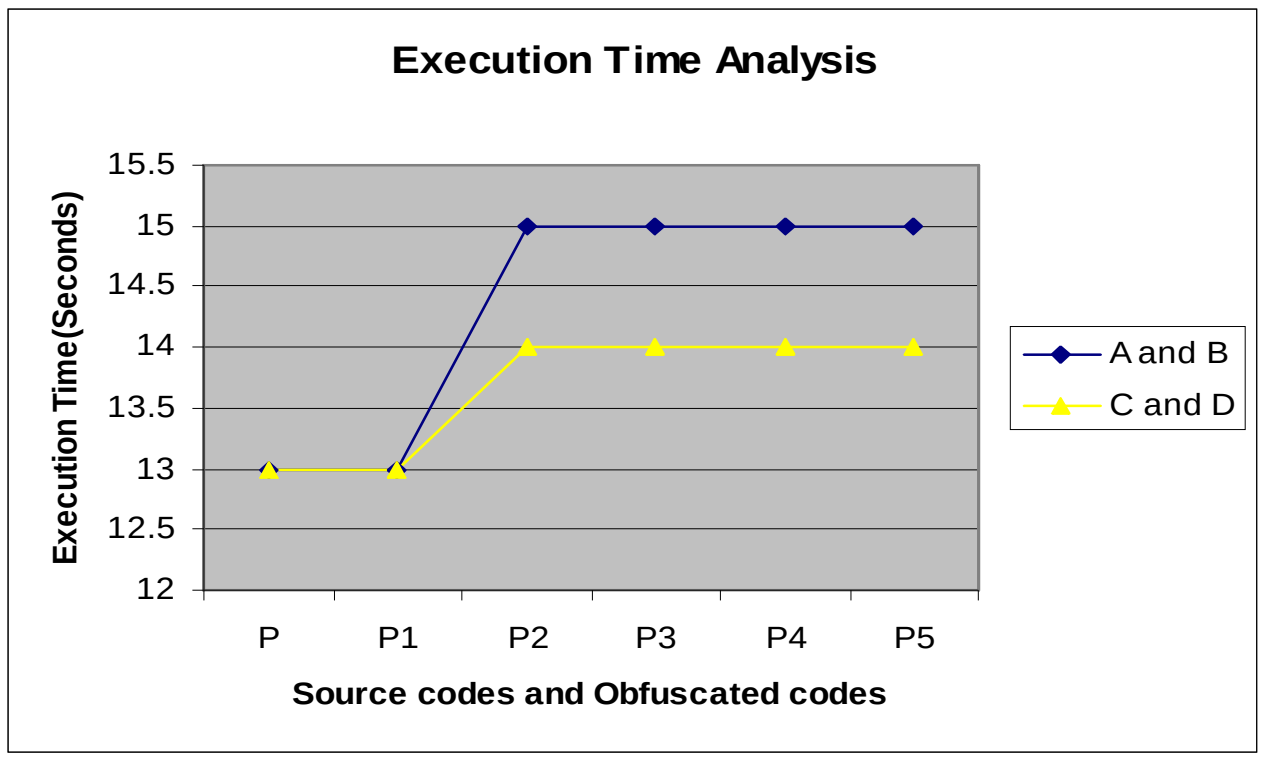

The above graph shows that there is not a considerable variation between execution times of the original code and obfuscated codes. The storage cost of the obfuscated files is measured in terms of file size.

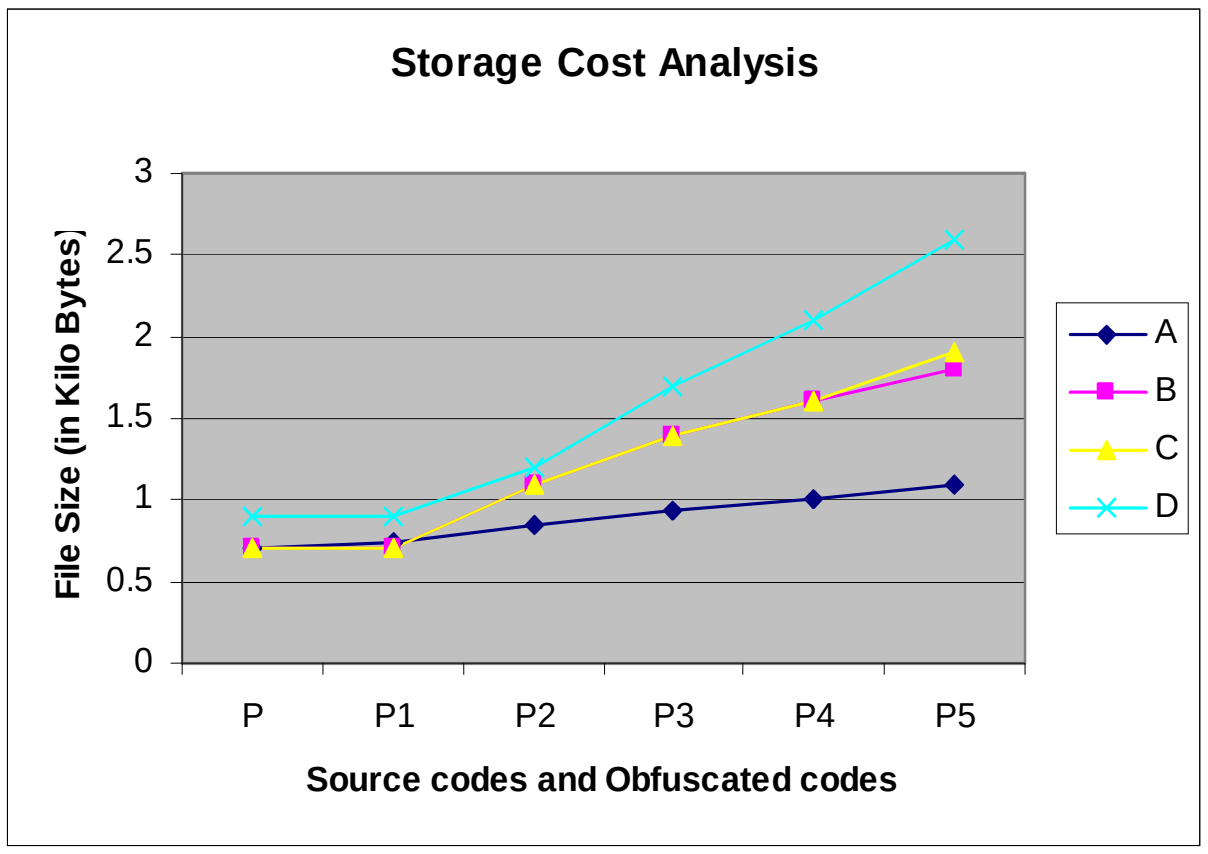

The graph shows that for code D, the file size grows more for obfuscated versions. The execution time is analyzed for Case1 program with random indices. The code for the following algorithm is denoted by 'E' and the obfuscated versions using SplitArray, FoldedArray and FlattenedArray are denoted by F, G and H respectively.

*For n>0, Set n different elements to array 'A' of size 100000*
`for (i : 0 .... n-1){`
*Generate a random number say 'num < n*
***Access** A[num]}*

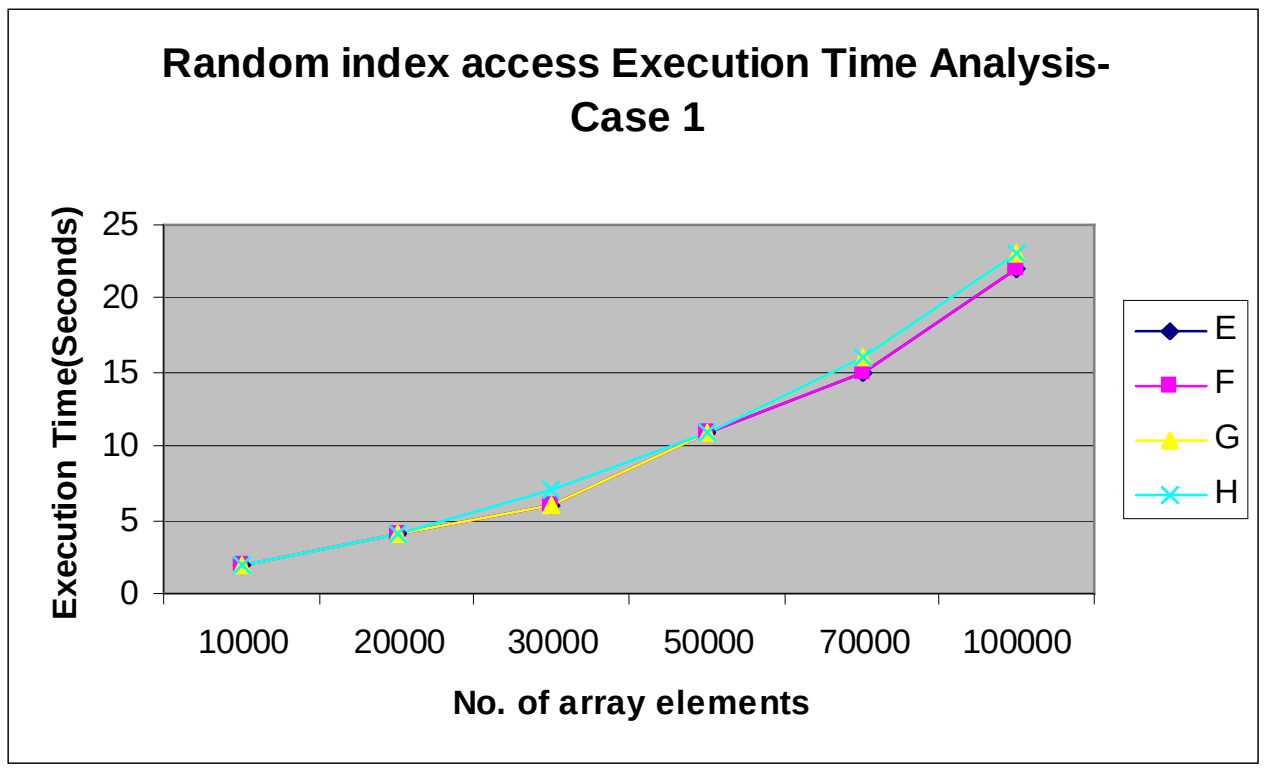

The graph shows no major variation for the execution times. For Case1, the execution times for 100000 elements are analyzed for codes and obfuscated versions of E, F, G and H.

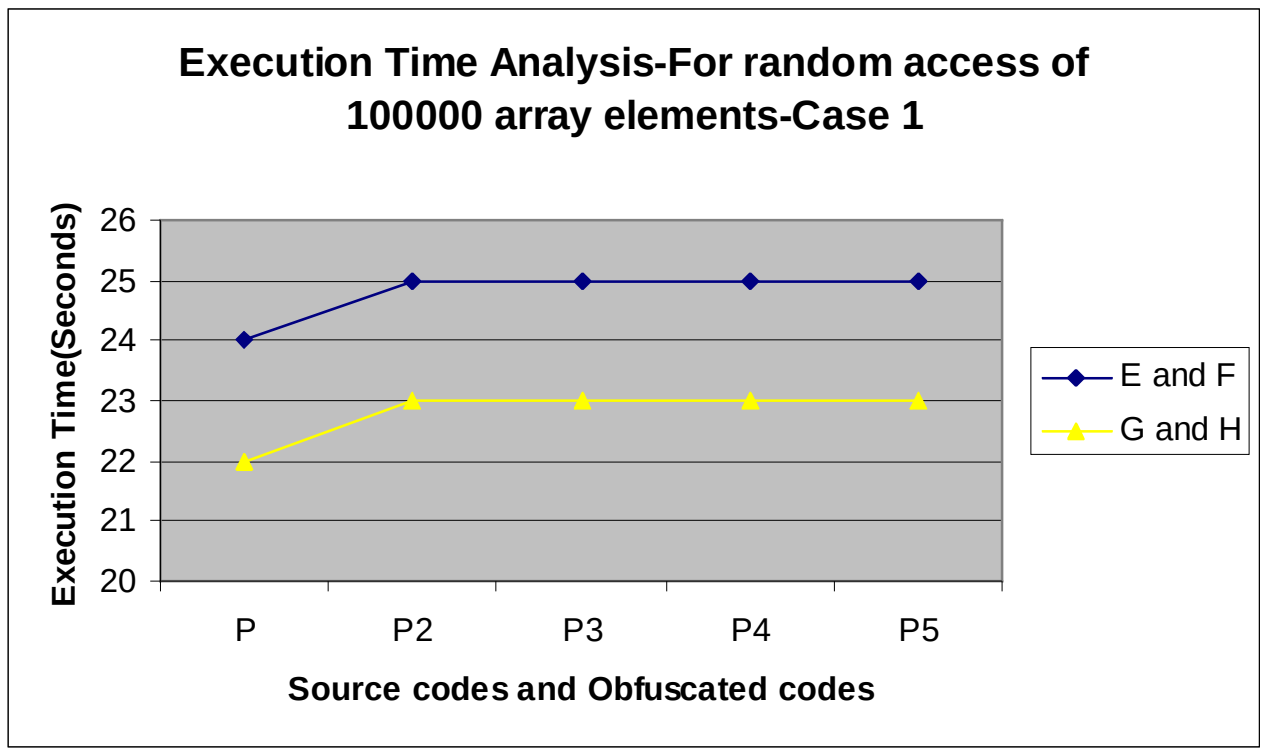

In the graph, considerable variation does not appear for execution times of E,F and G,H. Again, the execution time is examined for the following code of Case 2 with random indices, denoted by 'I'.

*Initialize array 'A' of size 100000, to 0*
*Read n*
```
for (i : 0 .... n-1){
```
*Generate a random number say 'num'<n*
*if((num%2)==0)* ***Access*** *A[num]*
*else* ***Set*** *A[num] }*

The obfuscated versions are represented by J, K, L.

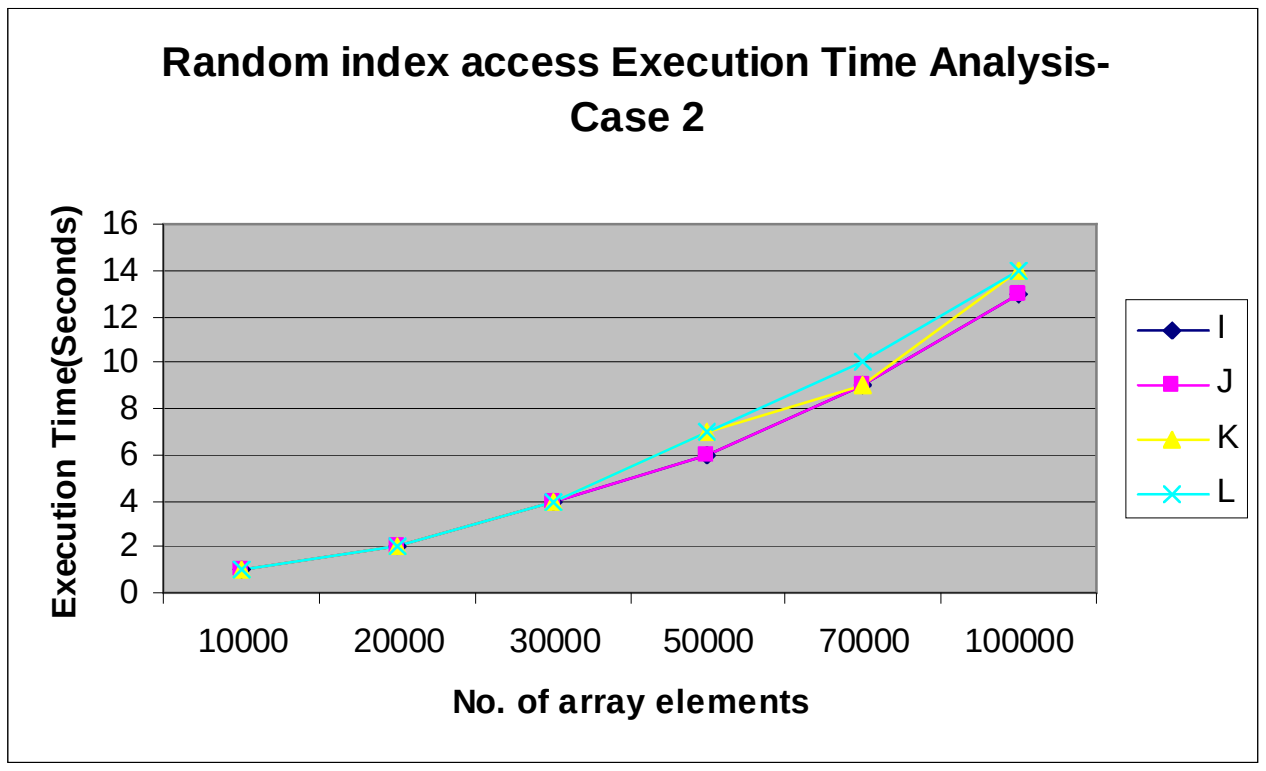

For case2, major variations are not occurring for execution times. For 100000 elements the execution times are examined for codes I,J,K,L and its obfuscated versions.

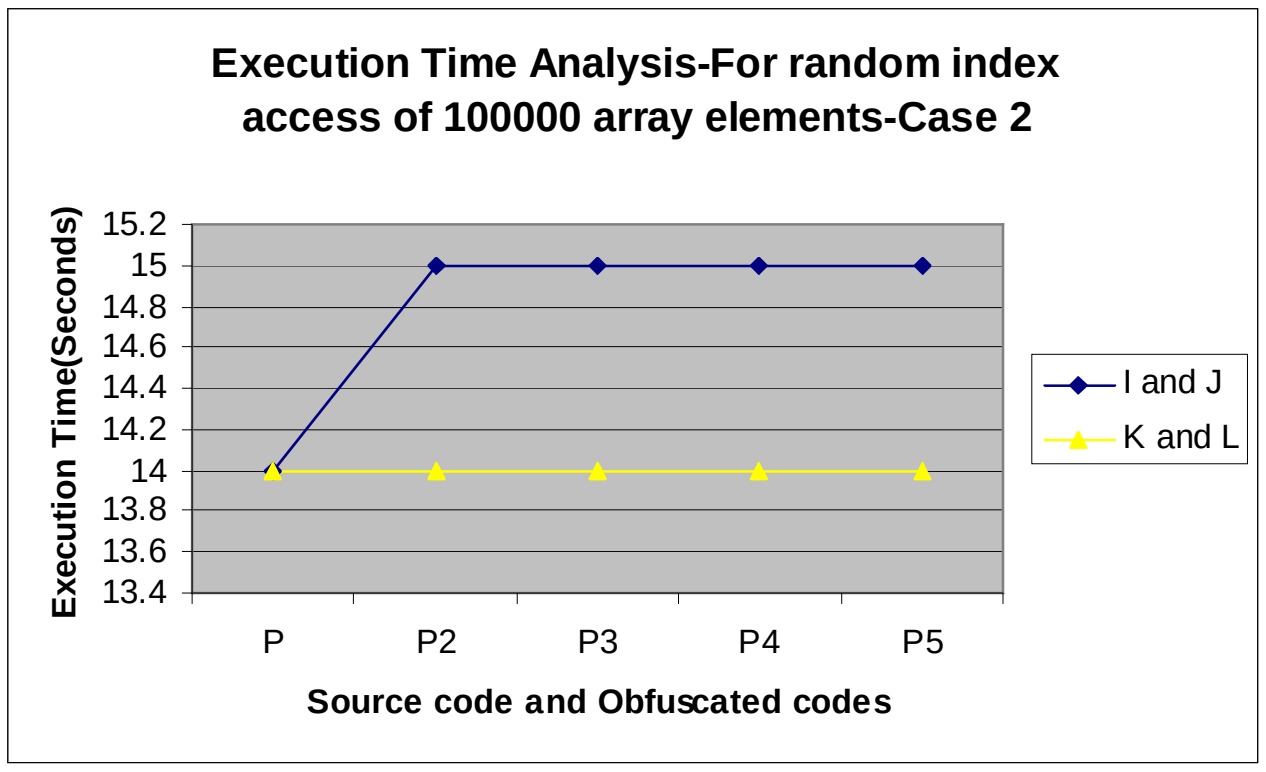

The graph illustrates maintenance of almost uniform execution times for codes I,J and K,L.

Due to the lack of commercial de-obfuscators in the market, this analysis is solely based on decompilation of code. The function call F(a,b) adds nearly 22 statements to the decompiled code and this call is crucial in adding obscurity to the code. In the next chart, we show the additional statements added in the code due to the recursive calls to F( ) functions due to multiple iterations of obfuscation. Let Fi denote the recursive call to F( ) upto depth i. Noting that the reverse engineering effort is the total time to understand entire code statements and is proportional to the number of statements. The decompiled codes of P2, P3, P4, P5, P6, P7 corresponding to A, B, C, D are analysed using the FrontEnd Plus v1.04 decompiler to find the number of F( ) calls in the decompiled codes.

| Decompiled Codes | Data Hiding Function representation | Data Hiding Function Calls | A | B | C | D |
|---|---|---|---|---|---|---|
| P2 | F1 | **F(a,b)** | **11** | **14** | **14** | **19** |
| P3 | F2 | **F((F(…)))** | **5** | **8** | **8** | **13** |
| P4 | F3 | **F((F((F(…)))))** | **5** | **8** | **8** | **13** |
| P5 | F4 | **F((F((F((F(….)))))))** | **5** | **8** | **8** | **13** |
| P6 | F5 | **F((F((F((F((F(…..)))))))))** | **5** | **8** | **8** | **13** |
| P7 | F6 | **F((F((F((F((F((F(….)))))))) ))** | **5** | **8** | **8** | **13** |

Table 1. Decompiled code Analysis

The reverse engineering effort added by the complex function calls for the decompiled codes is plotted in the following graph.

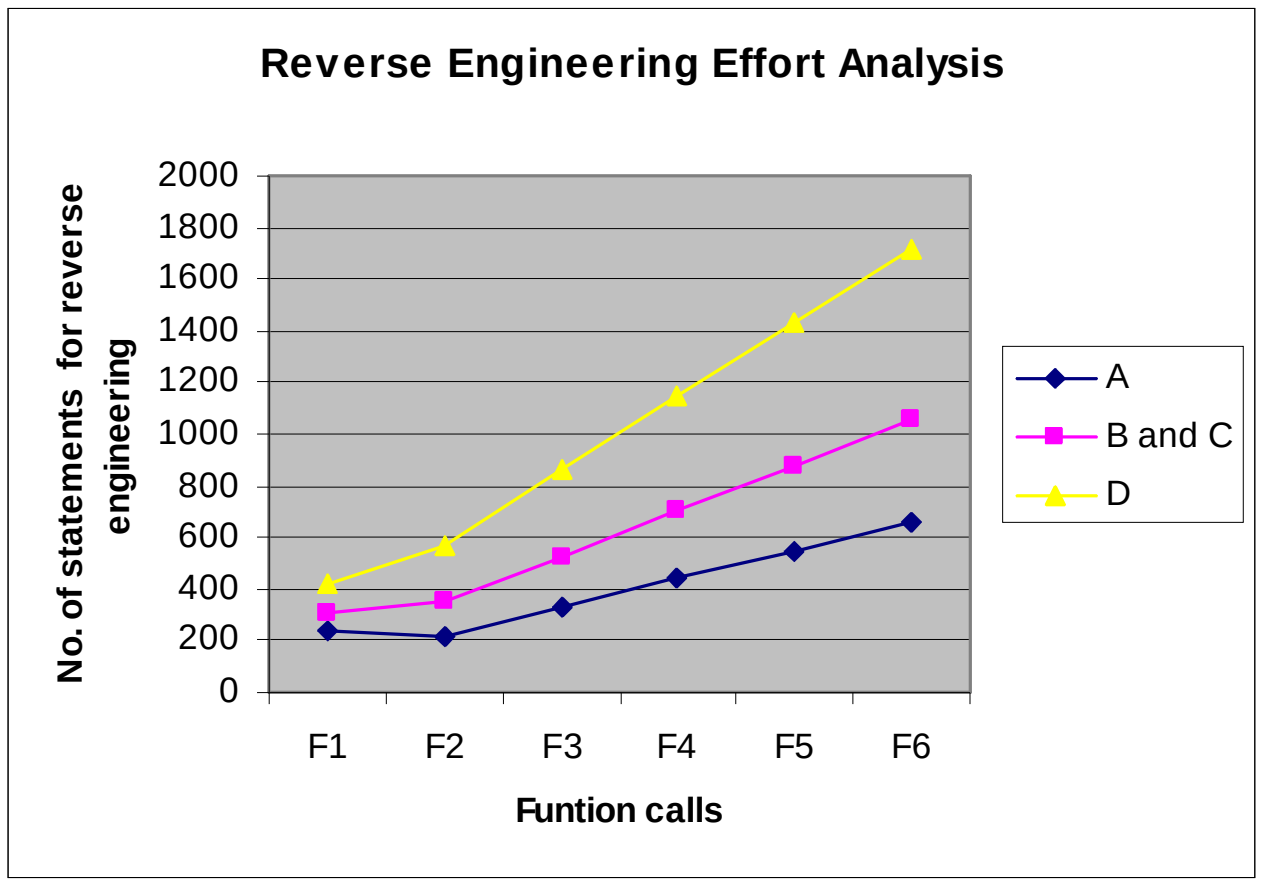

Considering the number of statements for reverse engineering, the reverse engineering effort grows considerably for codes B,C and D, but not significantly for A. We conclude that, further code obfuscation would add more effort in reverse engineering, without too much cost on execution time and storage.

## 4. Future Work

Generalized array splitting method [8] for generalizing array splitting and homomorphic obfuscations [20] for strengthening obfuscations are being incorporated in the tool.